\documentclass[aps,prd,superscriptaddress,eqsecnum,nofootinbib,showkeys,twocolumn,showpacs,preprintnumbers]{revtex4}
\usepackage{graphicx}
\usepackage{euscript,amssymb}
\usepackage{amsfonts}
\usepackage{amsmath}

\def\beq{\begin{equation}}
\def\eeq{\end{equation}}
\def\beqa{\begin{eqnarray}}
\def\eeqa{\end{eqnarray}}

\begin{document}

\title{Stochastic gravitational waves from a new type of modified Chaplygin gas}

\author{Mariam Bouhmadi-L\'{o}pez}
\email{mariam.bouhmadi@ist.utl.pt}
\affiliation{Centro Multidisciplinar de Astrof\'{\i}sica - CENTRA, 
Departamento de F\'{\i}sica, Instituto Superior T\'ecnico,
Av. Rovisco Pais 1, 1049-001 Lisboa, Portugal}
\author{Pedro Fraz\~ao}
\email{pedro.frazao@ist.utl.pt}
\affiliation{Centro Multidisciplinar de Astrof\'{\i}sica  - CENTRA,
Departamento de F\'{\i}sica, Instituto Superior T\'ecnico,
Av. Rovisco Pais 1, 1049-001 Lisboa, Portugal}
\author{Alfredo B. Henriques}
\email{alfredo@fisica.ist.utl.pt}
\affiliation{Centro Multidisciplinar de Astrof\'{\i}sica  - CENTRA,
Departamento de F\'{\i}sica, Instituto Superior T\'ecnico,
Av. Rovisco Pais 1, 1049-001 Lisboa, Portugal}

\begin{abstract}
We propose a new scenario for the early universe where there is a
smooth transition between an early de Sitter-like phase and a radiation dominated era.
In this model, the matter content is modelled  by a new type of generalised Chaplygin gas \cite{chaplygin} for the early universe, 
with an underlying scalar field description.
We study the gravitational waves generated by the quantum fluctuations. In particular, we calculate the gravitational
wave power spectrum, as it would be measured today, following  the method
of the Bogoliubov coefficients. We show that the high frequencies
region of the spectrum depends strongly  on one of the parameters of the
 model. On the other hand, we use the number of e-folds, along with the power spectra
 and spectral index of the scalar perturbations, to constrain the  model observationally.

\end{abstract}

\date{\today}


\maketitle

\section{Introduction}\label{sec1}

The inflationary paradigm is the most consistent scenario to explain the origin of the large 
scale structure (LSS) of the universe, as well as the anisotropies in the cosmic microwave background radiation (CMBR) \cite{LiddleLyth,Kinney:2009vz}. 
Originally, an early inflationary era in the universe  was invoked as a mechanism to solve several
shortcomings of the big bang theory 
\cite{inflation}. Afterwards, it was realised  that an early inflationary era of the universe 
generates density perturbations as seeds for the structure we see nowadays. In addition, a background of stochastic primordial gravitational waves (GW) is also predicted and it is  originated from the vacuum fluctuations.
The latter (GW) if ever detected in the future would provide an amusing source of information about the early universe. On the other hand, 
and most importantly, the predictions of the inflationary theory have been corroborated by several cosmological observations like the recent WMAP5 data \cite{WMAP}.

In this letter we propose a phenomenological model for the inflationary era and the subsequent radiation
 dominated epoch of the universe. In particular, we suggest a way to extend the framework of the generalised
 Chaplygin gas (GCG) to the first stages of the universe\footnote{An alternative inflationary model inspired on the Chaplygin gas was proposed in Ref.~\cite{Bertolami:2006zg} where only the inflationary era is accounted for.}. The GCG has attracted a lot of attention over the last years  \cite{chaplygin,chaplygin2,gorini,Barreiro:2008pn,GCGsingularity}. It was initially introduced  in cosmology as a mean to unify the dark sectors of the universe, 
i.e. dark energy and dark matter, and has been contrasted with several cosmological observations \cite{Barreiro:2008pn}.
On the other hand, it has been as well shown that some GCG models can be an excellent framework to analyse dark energy related singularities \cite{chaplygin2,GCGsingularity}.

In this work, we investigate the 
possible imprints in the power spectrum of the stochastic background of GW, for the model 
we propose for the transition from the inflationary era to the radiation dominated epoch\footnote{In Ref.~\cite{{Fabris:2004dp}}, the spectrum of GW for a universe whose dark energy component corresponds to a Chaplygin gas was analysed. This scenario is different from ours.}. This transition is
far from being well known and it can be of some interest to explore the signatures associated with it, 
shedding light on the inflationary model behind the accelerated expansion in the early universe.

We calculate the GW production using the Bogoliubov coefficients which obey a set of differential equations \cite{parker,Starobinsky,allen,mendes-henriques-moorhouse}. This method has advantages over the frequently used
sudden transition approximation, because, associated with it, there is always an 
overproduction of gravitons of large
frequencies, 
which is avoided in a natural way by the use of the continuous Bogoliubov
coefficients \cite{mendes-henriques-moorhouse}. This is also a
very practical method to calculate the full spectrum, from the
very low frequencies corresponding to the present cosmological
horizon $10^{-17}$Hz, till those large, kHz up to GHz, frequencies associated with the
transition between the inflationary  and the radiation-dominated universe.

The present letter is organised as follows. In section \ref{sec2}  we present our  
model based on a \textit{new type of generalised Chaplygin gas for the early universe}. This fluid can be as well 
represented by a minimally coupled scalar field as shown in section \ref{sec3}. 
 In section \ref{sec4}, we constrain our model observationally. In section \ref{sec5}, we summarise the  methodology used to obtain 
the spectrum of the stochastic GW which is based on the Bogoliubov coefficients. 
We present in section \ref{sec:NumericalSimulations} the spectrum of the GW  
for the model introduced in sections \ref{sec2} and \ref{sec3}. Finally, in section \ref{sec7} we present our conclusions.

\section{A modified Chaplygin gas for the early universe}\label{sec2}

We consider a perfect fluid whose energy density, $\rho$,  scales with the scale factor, $a$,  as
\beq
\rho=\left(A+\frac{B}{a^{4(1+\alpha)}}\right)^{\frac{1}{1+\alpha}},
\label{rho}
\eeq
where $A,B$ and $\alpha$ are constants. This energy density interpolates between a positive cosmological
 constant at small scale factors and a radiation fluid at large scale factors as long $1+\alpha$ is negative  and $A,B$ are positive. 

By assuming the conservation of the energy density of such a fluid (cf. Eq.~(\ref{rho})), 
which is a direct consequence of the Bianchi identity, the equation of state satisfied by its pressure, $P$, and energy density, $\rho$, reads
\beq
P=\frac13\rho -\frac43\frac{A}{\rho^{\alpha}}.
\label{equationstate}
\eeq
Before moving on a few words on the physics behind such a fluid are in order: like a generalised Chaplygin gas, whose equation of state reads  $P=-A/\rho^{\alpha}$, a fluid with the equation of state (\ref{equationstate}) could be formally related to an effective description of a complex scalar field whose action can be written as a generalised Born-Infled action \cite{chaplygin}, corresponding to a \textit{perturbed} d-brane in a $(d+1,1)$ space-time. For the propose of this work it is sufficient to describe such a matter content through the equation of state  (\ref{equationstate}) or through a scalar field as it is shown on the next section.

The equation of state (\ref{equationstate}) corresponds to a mixture of a radiation fluid and a generalised Chaplygin gas
 \cite{chaplygin,chaplygin2} although, it is only the  total energy density $\rho$ which is conserved. 
This type of equation of state has been previously introduced in the context of dark energy models 
\cite{Benaoum:2002zs,Gonzalez-Diaz:2002hr,Debnath:2004cd,Chimento:2005au} (see also \cite{odintsov,Barrow:1988yc,Nunes:2000yc}).  Nevertheless, we would like to stress that it has never been analysed in  the context of 
 the early evolution of the  universe. More precisely, the equation of state 
(\ref{equationstate}) has been analysed in Ref.~\cite{Benaoum:2002zs} for $0<\alpha<1$; i.e. a fluid interpolating between   
a radiation at early time and a cosmological constant at late time. Here we will be analysing a completely 
different range of the space parameter where $\alpha<-1$.

For $\alpha<-1$ the energy density (\ref{rho}) can be approximated as 
\beqa
  \rho&\simeq& A^{\frac{1}{1+\alpha}} \quad \,\,a\ll a_\star, \label{rhoinf}\\
  \rho&\simeq& \frac{B^{\frac{1}{1+\alpha}}}{a^4} \quad a\gg a_\star,\label{rhorad}
\eeqa
where
\beq
a_{\star}=\left(\frac{B}{A}\right)^{\frac{1}{4(1+\alpha)}}.
\eeq
Therefore, a homogeneous and isotropic universe  with this matter content (cf. Eqs.~(\ref{rho}) and (\ref{equationstate})) 
undergoes a primordial inflationary era; initially described by a de Sitter (dS) expansion (see Eq.~(\ref{rhoinf})). 
Then, the universe keeps inflating until the scale factor reaches $a_\star$ where inflation ceases; i.e. $\rho+3P=0$. 
Finally, it enters a radiation dominated epoch (see Eq.~(\ref{rhorad})). The transition from the inflationary 
era to the radiation dominated phase takes place smoothly.

The Friedmann equation 
\beq
H^2=\frac{\kappa^2}{3}{\rho},
\label{friedmann} 
\eeq
where $\kappa^2=8\pi\rm{G}$ and $\rho$ is defined in Eq.~(\ref{rho}),  can be 
integrated analytically\footnote{This Friedmann equation
can be integrated in two steps by (i) performing it for $0<1+\alpha$ \cite{W} and (ii) 
performing an analytical continuation of the hypergeometric function to values of $\alpha$ such that $1+\alpha <0$ \cite{X}.\label{footnote1}} \cite{W,X}

\beqa
&&\frac{\kappa a_{\star}}{\sqrt3}A^{\frac{1}{2(1+\alpha)}}(\eta-\eta_\star)=2^r\,F\left(-r,1;1-\frac{r}{2};\frac12\right)
\nonumber\\
&\,\,\,\,&-y^{-\frac{r}{2}}(1+y)^r\,F\left(-r,1;1-\frac{r}{2};\frac{y}{1+y}\right).
\label{aeta}
\eeqa
In the previous expression  $\eta$ is the conformal time, $\eta_\star$ is a constant, F(a,b;c;y) is a hypergeometric function\footnote{A hypergeometric series $\textrm{F}(b,c;d;x)$, also called a hypergeometric
function, converges at any value $x$ such that $|x|\leq 1$,
whenever $b+c-d<0$. However, if  $0 \leq  b+c-d < 1$ the series
does not converge at $x=1$. In addition, if  $1 \leq b+c-d$, the
hypergeometric function blows up at $|x|=1$
\cite{W}.\label{series}} and   
\beq
y=\left(\frac{a}{a_{\star}}\right)^{-4(1+\alpha)},\quad r=-\frac{1}{2(1+\alpha)}\,.
\label{definitionyr}
\eeq
At the conformal time $\eta_\star$ the universe exits the inflationary era as $a=a_\star$. 

The primordial universe starts its dS expansion at $\eta\rightarrow -\infty$ where $a\ll a_\star$. On the other hand, the universe gets radiation-dominated at $\eta\rightarrow \infty$ where $a_\star\ll a$. At this respect,  we note that \cite{X}
\beq
F\left(-r,1;1-\frac{r}{2};1\right)=-1.
\eeq
By choosing
\beq
\frac{\kappa a_{\star}}{\sqrt3}A^{\frac{1}{2(1+\alpha)}}\eta_{\star}=-2^r\,F\left(-r,1;1-\frac{r}{2};\frac12\right),
\eeq
the relation between the conformal time, $\eta$, and the scale factor given in Eq.~(\ref{aeta}) can be simplified to
\beqa
\kappa a_{\star}A^{\frac{1}{2(1+\alpha)}}\eta&=&-\sqrt{3}y^{-\frac{r}{2}}(1+y)^r\,\times\\
&\,\,&\times\,\,
F\left(-r,1;1-\frac{r}{2};\frac{y}{1+y}\right)\nonumber.
\eeqa
We notice that this redefinition of $\eta$ does not modify the fact that in this model the universe has an infinite past (in terms of the conformal time), where it is asymptotically dS, and an infinite future (in terms again of the conformal time), where it is radiation-dominated. 

The whole evolution of the universe can be described by the modified Chaplygin gas (\ref{equationstate}) at early-time and by the $\Lambda$CDM model at late-time; i.e.
\begin{eqnarray} \rho = \left\{ \begin{array}{lll}
          & \left(A+\frac{B}{a^{4(1+\alpha)}}\right)^{\frac{1}{1+\alpha}} &\,\,\mbox{early-time} \\
          & \,\, & \\                                                     
          & \rho_{r0}\left(\frac{a_0}{a}\right)^4+\rho_{m0}\left(\frac{a_0}{a}\right)^3+\rho_\Lambda & \,\,\mbox{late-time }\end{array} \right.  \label{energydensities}\end{eqnarray}
where $a_0$ is the current scale factor, $\rho_{r0}$, $\rho_{m0}$ are $\rho_\Lambda$ are the current energy densities corresponding to radiation, matter and  the cosmological constant, respectively. On the other hand, at the radiation dominated epoch the energy densities (\ref{energydensities}) are equal implying 
\begin{equation}
B=\left(\rho_{r0}a_0^4\right)^{1+\alpha}\label{Bdef}.
\end{equation}
Finally, we notice that $A$ is related to the scale of inflation (see Eq.~(\ref{rhoinf})).

\section{The Inflationary Dynamics}\label{sec3}

The inflationary dynamics of the model presented in section \ref{sec2} is better described through a minimally coupled scalar field, $\phi$, with a potential, $V(\phi)$, whose energy density and pressure read
\beqa
\rho_\phi=\frac{\phi'^2}{2\,a^2}+V(\phi)\qquad , \qquad P_\phi=\frac{\phi'^2}{2\,a^2}-V(\phi).
\eeqa
In the previous equations the prime stands for derivative respect to the conformal time.

\begin{figure}[h]
  \includegraphics[width=5.5cm]{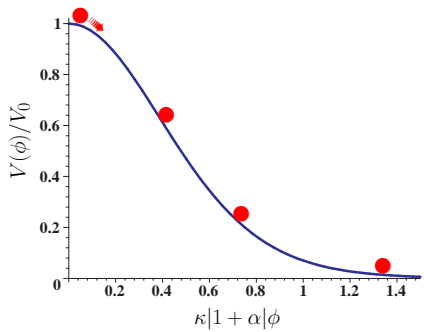}
	\caption{Plot of the potential (\ref{scalarFieldPotential}) as a function of the scalar $\phi$ given in Eq.~(\ref{phia}).}
	\label{V1}
\end{figure}

In a Friedmann-Lema\^itre-Robertson-Walker (FLRW) universe, the scalar field, $\phi$, is equivalent to the modified Chaplygin 
gas introduced in Eqs.~(\ref{rho}) and  (\ref{equationstate}), 
as long as $\rho_\phi=\rho$ and $P_\phi=P$. The last two equalities imply\footnote{A similar potential for a different equation of state was suggested in Ref.~\cite{Bernardini:2007hm}.}
\beqa
V(\phi)&=&\frac{V_0}{3}\left[\cosh^{\frac{2}{1+\alpha}} \Big(\kappa|1+\alpha|\,\phi\Big)\right.\nonumber\\
  &\,&\,\,\,\,\,\,\,\,\left.+2\;\cosh^{-\frac{2\,\alpha}{1+\alpha}} \Big(\kappa|1+\alpha|\,\phi\Big)\right],
\label{scalarFieldPotential}
\eeqa
where 
\beq
V_0=A^{\frac{1}{1+\alpha}}\label{V0}.
\eeq
In addition,  $\phi$ can also be determined analytically in terms of  the scale factor
\beq
\phi(a)=\pm\frac{1}{\kappa|1+\alpha|}\mathrm{arcsinh} \left(\sqrt{\frac{B}{A}}\;a^{-2(1+\alpha)}\right).\label{phia}
\eeq
For simplicity, we will restrict to the solution (\ref{phia}) with + as our results does not depend on which of the two signs we choose. 
The universe starts its inflationary phase in a quasi-dS state where the scalar field is sitting close the top of the potential; i.e. at $\phi\sim 0$ (see Fig.~\ref{V1}). 
Then it starts rolling down the potential until it exits the inflationary era when $\phi(a_\star)=\phi_\star$; i.e. 
\beq
\phi_\star=\frac{\ln(1+\sqrt{2})}{\kappa|1+\alpha|}.
\eeq
Finally, the radiation dominated phase starts for large values of the scalar field, i.e. $\phi_\star\ll\phi$. For latter convenience, we derive as well  $\rho_\phi$ in terms of the scalar field
\beqa
  \rho_\phi=V_0\;\text{cosh}^{\frac{2}{1+\alpha}}\left(\kappa|1+\alpha|\phi\right).
\label{rhophia}
\eeqa

The number of e-folds of expansion since a given mode $k$ exits the horizon  ($k=aH$) during the inflationary era, at $\phi=\phi_c$, until the end of inflation can be approximated by 
\footnote{Here, we are making use of a slow roll approximation \cite{Lidsey97}; 
i.e. values of the scalar field such that $\epsilon, |\eta|\ll 1$ where $\epsilon$ and $\eta$ are the standard slow-roll parameters \cite{Lidsey97} (see also Eq.~(\ref{slparameter})). } 
\beq
N_c\thickapprox \kappa^2\int_{\phi_{\star}}^{\phi_c}V\left({\frac{dV}{d\phi}}\right)^{-1}\;d\,\phi
\eeq
Substituting the scalar field potential (\ref{scalarFieldPotential}) on the previous expression,  we obtain
\beqa
  N_c(\alpha,\phi_c)&=&\frac{1}{2\,(1+\alpha)(2\,\alpha-1)}\ln \Bigg\{\sinh\big(\kappa|1+\alpha|\phi_c\big)^{-3}\nonumber\\
&\times&  \left[\frac{1-4\,\alpha+\cosh\big( 2\kappa|1+\alpha|\phi_c\big)}{4(1-\alpha)}\right]^{1+\alpha}\Bigg\}
\label{NumberEFoldsModel}
\eeqa

\section{Observational Constraints}\label{sec4}

\begin{figure}[h]
\begin{center}
\includegraphics[width=6.5cm]{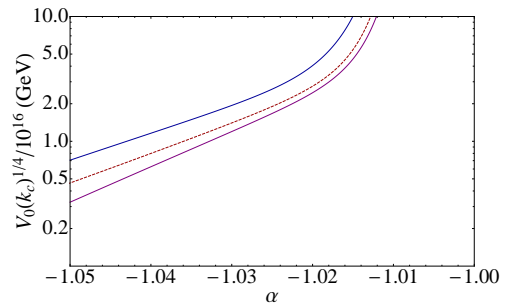}
\end{center}
\caption{Variation of the inflationary scale versus $\alpha$. Using the first method, i.e. the number of e-folds and $P_s$, we obtain the dashed-dotted and dashed lines corresponding to $N_{\rm c}=47$ and $N_{\rm c}=62$, respectively. Using the second method; i.e. $P_s$ at the pivot scale $k_c=0.05\;\mathrm{Mpc^{-1}}$, we obtain the solid line.}
\label{fig:InflationaryScale}
\end{figure}

We constrain our model through CMBR/LSS measurements, using the values \mbox{$P_s(k_c)=(2.45\pm 0.23)\times 10^{-9}$} and  
$n_s=1.0\pm 0.1$ for the power spectrum and spectral index of the scalar perturbations, respectively,
 evaluated at the pivot wave number $k_c=0.05\;\mathrm{Mpc^{-1}}$, that corresponds to CMBR/LSS scales that exited the horizon during inflation \cite{Lidsey97}.
To first order in the slow-roll approximation, the amplitude of the power spectrum for density perturbations is given by \cite{Lidsey97}
\beqa
     P_{s}(k) &\approx& \frac{\kappa^6}{12\pi^2}\frac{V^3}{V_\phi^{2}} \Bigg|_{k=aH}, \label{eq:scalar amp}
\eeqa
where $V_\phi\equiv {dV\over d\phi}$. Notice that $P_{s}$ is proportional to $V_0$ and therefore the power spectrum  is extremely useful to constrain the energy scale of inflation (cf. Eqs.~(\ref{rhoinf}), (\ref{scalarFieldPotential}) and (\ref{V0})). We will follow two different methods to constrain $V_0$:

\begin{itemize}

\item Method 1: As a first step to constrain our model, we use the bounds on $N_c$, i.e. the number of e-folds since a given mode exits the horizon at $\phi_c$ until the end of inflation. We consider as a conservative range $47\leq N_c\leq 62$ \cite{Lidsey97}. Then for a given $N_c$, on the allowed range, and for a fixed $\alpha$, we obtain the corresponding value $\phi_c$ (cf. Eq.~(\ref{NumberEFoldsModel})). Finally, using the pair $(\alpha,\phi_c)$ in Eq.~(\ref{eq:scalar amp}) with \mbox{$P_s(k_c)=(2.45\pm 0.23)\times 10^{-9}$}, we deduce the scale of inflation; i.e. $V_0$. This is precisely, what is shown on the dashed-dotted  and dotted lines of Fig.~\ref{fig:InflationaryScale}.

\item Method 2: A tighter constraint on $V_0$ is obtained by  evaluating $P_s$ at the pivot scale $k_c$ when the mode exits the horizon; i.e. $k_c= aH$, during the inflationary era. Using Eqs.~(\ref{friedmann}), (\ref{phia}) and (\ref{rhophia}), this scale reads  
\beqa
  &&\hspace{-0.6cm}k_c=\frac{2\pi}{\sqrt{3}} \kappa \rho_{r0}^{1/4}V_0^{1/4}\;a_0\times\nonumber\\
  &&\times\Big( \text{cosh}\big(\kappa|1+\alpha|\phi_c\big)\;\text{coth}\big(\kappa|1+\alpha|\phi_c\big)\Big)^{\frac{1}{2(1+\alpha)}}.
\eeqa
Finally, by combining the previous equation and Eq.~(\ref{eq:scalar amp}), we deduce the corresponding scale $V_0$ for a given parameter $\alpha$. The results are shown on the solid line in Fig.~\ref{fig:InflationaryScale}.

\end{itemize}

\begin{figure}[h]
  \centering
  \includegraphics[width=7.5cm]{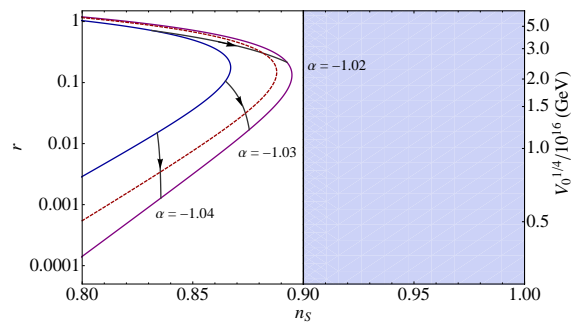}
	\caption{Constraints on the $n_s-r$ parameter space. Using the first method; i.e. i.e. the number of e-folds and $P_s$, we obtain the dashed-dotted and dashed lines corresponding   to $N_{\rm c}=47$  and $N_{\rm c}=62$, respectively. Using the second method; i.e. $P_s$ at the pivot scale $k_c=0.05\;\mathrm{Mpc^{-1}}$, we obtain the solid line. The black lines are for
a constant $\alpha$-value, parameterised by an increase in the number of e-folds.}
	\label{nSrParamenterSpace}
\end{figure}

We next consider the tensor to scalar ratio 
\beqa
  r \equiv \frac{P_t(k)}{P_s(k)} = 16\,\epsilon\,, 
\eeqa
where $P_{t}(k)$ is the tensorial power spectrum \cite{Lidsey97}. We analyse the variation of this ratio against the spectral index of the scalar perturbations  $n_s=1-6\,\epsilon+2\,\eta$. In the previous expressions the slow-roll parameters read
\begin{equation} 
  \epsilon(\alpha,\phi_c)=\frac{1}{2\kappa^2}\left(\frac{V_\phi}{V}\right)^{2};\quad \eta(\alpha,\phi_c)=\frac{1}{\kappa^2}\frac{V_{\phi\phi}}{V}.\label{slparameter}
\end{equation}
We follow the two approaches enumerated before to evaluate $r$ and $n_s$. Our results are shown on the $n_s-r$ parameter space in Fig.~\ref{nSrParamenterSpace}. We see that the spectrum is slightly more red than preferred by the observation. However, Fig.~\ref{nSrParamenterSpace} shows that $r<1$ is in agreement with the observation \cite{Lidsey97}. Our best fit value corresponds to $\alpha=-1.024$; i.e. an energy scale for inflation about $10^{16}$ GeV. It is worthy to notice that for a standard Chaplygin gas accommodated to the late-time expansion of the universe,  negative values of $\alpha$ are allowed by Type Ia Supernova\footnote{We thank O. Bertolami for pointing out this to us.} \cite{Bertolami:2004ic}.

\section{Gravitational Wave Spectrum}\label{sec5}

The stochastic gravitational wave (GW) spectrum, generated during the expansion of the universe, can be calculated using 
the method of continuous Bogoliubov coefficients, using the methods first developped by L. Parker \cite{parker}  and A. A. Starobinsky \cite{Starobinsky},  and later applied in \cite{mendes-henriques-moorhouse,abh1}. Indeed, as the universe evolves, the annihilation  and creation  operators change with time.

We can write these operators in terms of time-fixed annihilation 
 and creation operators through a Bogoliubov transformation, involving the Bogoliubov coefficients $\alpha$ and $\beta$,
which must satisfy the constraint $|\alpha|^2-|\beta|^2=1$ \cite{parker,mendes-henriques-moorhouse}.

It can be shown that the
dimensionless relative logarithmic energy spectrum of the GWs, $\Omega_{\rm GW}$, 
at the present time $\eta_0$, is related with $\beta$ \cite{abh1}:
\beqa
\Omega_{\mbox{\scriptsize GW}} &\equiv&\frac{1}{\rho_{\mbox{\rm c}}}
\frac{d\rho_{\mbox{\scriptsize GW}}}{d \ln \omega} 
= \frac{\hbar \kappa^2}{3\pi^2 c^5 H^2(\eta_0)} \omega^4 \beta^2, \label{8}
\eeqa
where $\rho_{\rm GW}$ is the energy density of GWs and  $\omega$  the respective
 angular frequency; $\rho _{\rm c}$ and $H$ are the critical density of the
universe and Hubble parameter, respectively, evaluated at the present time.

Therefore, the evolution of the Bogoliubov coefficient $\beta$, since early times, 
gives us the power spectrum at the present time. 
The same coefficient can be written in terms of two continuous functions of time, $X$ and $Y$, in the form
 $|\beta|^2=(X-Y)^2/4$. These new functions are determined by the set of equations  \cite{abh1}
\beqa
X^\prime &=& -i k Y, \label{6a} \\
Y^\prime &=& -\frac{i}{k} \left( k^{2}-\frac{a^{\prime \prime }}{a} \right) X.
\label{6}
\eeqa

The Friedmann equation (\ref{friedmann}) and the conservation of the energy momentum tensor
 imply $a^{\prime \prime}/a\propto \rho-3\,p$. Therefore,
the power spectrum depends on the matter content of the universe (see for example Eq. (\ref{6})). 
Consequently, the GWs power spectrum can provide us with important information
about the universe evolution and its matter content, as long as the latter leaves a significant imprint on the spectrum.

As our model describes a transition between the inflationary period and the
radiation-dominated era, the integration  
must begin at the inflationary period, where we have a constant energy density.  
Hence, it is necessary to specify, at this inflationary
period, the initial conditions for $X(\eta)$ and $Y(\eta)$.

To solve the set of differential equations (\ref{6a}) and
(\ref{6}), for our model, we must use numerical methods, 
but, since it admits an exact analytical solution
for the case of a dS universe \cite{abh1}, i.e. similar to our model at early-time (see Eq.~(\ref{rhoinf})), we can specify
appropriate initial conditions for $X(\eta)$ and $Y(\eta)$, at an appropriate scale of inflation. 
In this case, Eqs.~(\ref{6a}) and (\ref{6}) yield a solution for a dS expansion that takes the form  \cite{abh1}
\beqa 
 X(\eta_{i})&=&\left(1+i\frac{a_iH}{k} \right) e^{i\frac{k}{a_iH}}, \label{InCondX}\\
\hspace{-3mm} Y(\eta_{i})&=& \left( 1 + i\frac{ a_iH}{k}-\frac{a_i^2H^2}{k^2} \right) e^{i\frac{k}{a_iH}}. \label{IniCondY}
\eeqa
where $a_i$ is a scale factor during the dS phase. 
We use these expressions as initial conditions for the numerical integration of Eqs.~(\ref{6a})-(\ref{6}). 

The other piece of information we need to solve Eqs.~(\ref{6a})-(\ref{6}) is  the whole evolution of the universe:  
\begin{eqnarray} 
\frac{a^{\prime \prime}}{a}= \left\{ \begin{array}{lll}
          & \frac{2}{3}\kappa^2a^2\,A\left(A+\frac{B}{a^{4(1+\alpha)}}\right)^{-\frac{\alpha}{1+\alpha}} &\,\,\mbox{early-time}\\
          & \,\, & \\                                                     
          & \frac{\kappa^2}{6}a^2\left[\rho_{m0}\left(\frac{a_0}{a}\right)^3+4\rho_\Lambda\right] & \,\,\mbox{late-time }
\end{array} \right.\, 
 \end{eqnarray}
where the transition between the two epochs is determined by the moment when ${a^{\prime \prime}}/{a}$ (for both periods)   is  approximately equal at $a=a_{\rm int}$.
We use the values for the energy densities parameters given by WMAP5
\cite{WMAP5data}:
$\Omega_m=0.291\pm0.014$ and $\Omega_\Lambda=0.709\pm 0.014$, through the usual definition $\Omega=\frac{8\,\pi}{3H_0^{\phantom{1}2}}\rho$, where $H_0=71.3\;\mathrm{km/s/Mpc}$.

\begin{figure}[h]
  \centering
  \includegraphics[width=7cm]{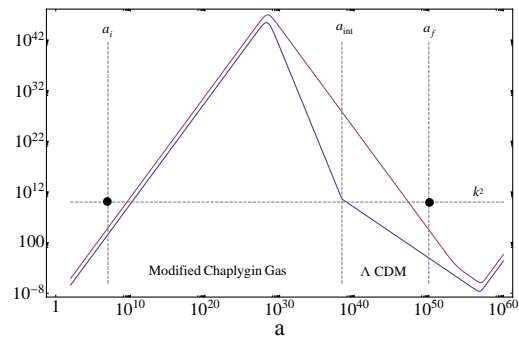}
	\caption{A scheme of the integration method used in the numerical simulations for a given mode $k$. The solid line refers to $\frac{a''}{a}$ and the dashed-dotted
one to $k_H^2=(a H)^2$.}
	\label{IntegrationMethod}
\end{figure}

For a given mode $k$, the integration is done in terms of the scale factor $a$. We start at $a \lesssim a_i$, just before the mode 
exits the horizon; i.e. $k<k_H$ where $k_H=aH$, until $a \gtrsim a_f$, when the mode reenters the horizon; i.e. $k<k_H$  (c.f.  Fig.~\ref{IntegrationMethod}). The maximum value of $k^2$ corresponds to the maximum height of the potential $a''/a$ as the gravitons are produced only below the barrier $a''/a$ (see Fig.~\ref{IntegrationMethod})

To ensure the accuracy of the numerical 
results are not a numerical artifact
we use the condition on the Bogoliubov coefficients $|\alpha|^2-|\beta|^2=1$,  which in terms of the new variables $X,Y$ reads $2XY=1$ , as a constraint equation of the numerical solution.

\section{Numerical Simulations}\label{sec:NumericalSimulations}

\begin{figure}[h]
  \centering
  \includegraphics[width=7cm]{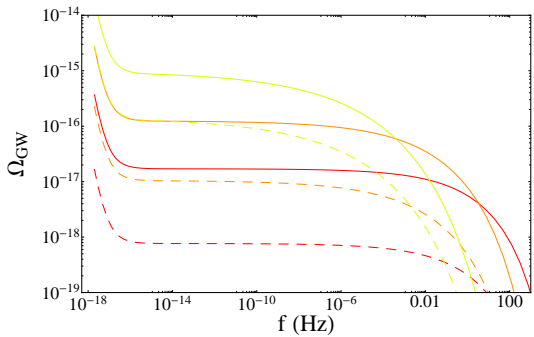}
  \caption{Gravitational wave spectra for the values $\alpha=-1.05\;(\text{black}),\;-1.04\;(\text{gray}), \text{ and } -1.03\;(\text{light gray})$,  
along with $N_c=47$ (solid) and $N_c=62$ (dashed). The spectrum suffers a downward shift with an increase  in $|\alpha|$, and, 
simultaneously, an increase in the high frequency region.}
  \label{AlphaVariation_CMB_LSS_1}
\end{figure}

\begin{figure}[h]
  \centering
  \includegraphics[width=7cm]{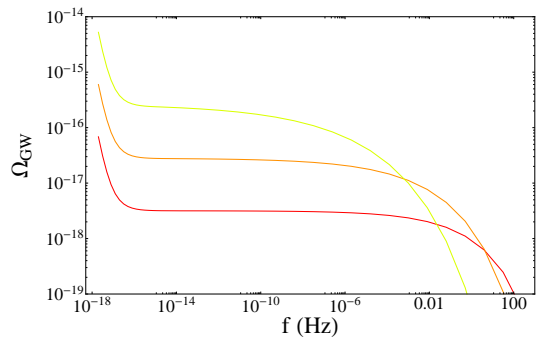}
  \caption{Gravitational wave spectra for the values $\alpha=-1.05\;(\text{black}),\;-1.04\;(\text{gray}), \text{ and } -1.03\;(\text{light gray})\;$. 
The spectrum suffers a downward shift with an increase in $|\alpha|$, and, 
simultaneously, an increase in the high frequency region.}
  \label{AlphaVariation_CMB_LSS_2}
\end{figure}

The GW spectra  resulting from the numerical integrations for the two different methods presented in section IV to constrain the model observationally are shown in Figs~\ref{AlphaVariation_CMB_LSS_1} and \ref{AlphaVariation_CMB_LSS_2}. In the former method, we use as a constraint equation the number of e-folds, $N_c$,  and in the latter the mode function, $k_c$. %
Therefore, as one would expect, the variation with the values of $\alpha$ is the same, independently of the method used to calculate the inflationary scale which gives us the constant $A$ of our model (see Eq.~(\ref{V0}) and Fig.~\ref{fig:InflationaryScale}). 
As we can see in Figs~\ref{AlphaVariation_CMB_LSS_1} and \ref{AlphaVariation_CMB_LSS_2}, the spectrum changes with $\alpha$, in particular it suffers simultaneously a vertical displacement and a variation in the high frequency region.

Furthermore, to understand the vertical displacement of the plateau in the intermediate region of the spectrum, it is necessary to consider the inflationary scale as a function of $\alpha$, plotted in Fig.~\ref{fig:InflationaryScale}.
The same figure shows that, for $\alpha$ constant, an increase in the number of e-folds from $N_c=47$ (dashed-dotted line) to $N_c=62$ (dotted line) results in a decrease in the inflationary scale $V_0$. 
Therefore, the decrease of the plateau height\footnote{An early dS phase results in a flat plateau in the spectra. The height of the plateau is an increasing function of the Hubble rate of the dS expansion; i.e. the energy scale of the dS inflationary era.}, which can be observed  in Fig.~\ref{AlphaVariation_CMB_LSS_1} by the difference between the solid lines ($N_c=47$) and the dashed lines ($N_c=62$) for a constant $\alpha$ (a given type of line), is easily interpreted as a decrease in the inflationary scale.

On the other hand, having showed that the vertical displacement of the spectra depends only on the value of the inflationary scale, we immediately conclude that the increase in the high frequency region, when $\alpha$ moves away from $-1$,  is a function of the parameter $\alpha$ only, as we can see for example in Fig.~\ref{AlphaVariation_CMB_LSS_2}.

The results presented above show that, in the gravitational-wave energy spectra for the unified inflationary and radiation eras consistent with CMBR/LSS constraints, and in particular the one by means of the mode function in Fig.~\ref{AlphaVariation_CMB_LSS_2},
present a variation in the spectrum that is a combination of the last effects: a vertical shift as
a result from the decrease in the inflationary scale; and an increase in the high frequency region that results from a increase in the $|\alpha|$-value. 

The last variation is the more important one because it can give us a new insight about the relation between the frequency limits of the GW spectrum and the model behind the inflationary  and the radiation eras.

Although the spectra presented on this section are smooth, reflecting the smoothness of the transition between the inflationary and the radiation dominated periods, characteristic of the present model, this is not always the case. For instances, when the system goes through a complex transition period, with changes in the equation of state, the structure of the spectrum is equally complex, particularly in the hundreds of MHz and GHz regime, allowing us to read a lot of information about the properties of the transition (see ref.~\cite{abh1}).

What about the reheating of the universe in our model? such a process can be modelled by coupling the scalar field to a radiation fluid, where the decay rate of the scalar field into the radiation fluid is governed by a phenomenological parameter. In addition, the tail of the potential (\ref{scalarFieldPotential}) is \textit{substituted} by a potential with a vanishing minimum. For the modified model the trend and the level of the spectra of the GW  would not be much different to the one presented on this section. Indeed, it is still the parameter $\alpha$ that characterises the spectra. In addition, the inclusion of a reheating process is computationally very much time consuming.

\section{Conclusions}\label{sec7}
\label{sec:Conclusions}

In this work, we investigate the production of GWs
in a new modified generalised Chaplygin gas for the early universe (cf. Secs \ref{sec2} and \ref{sec3}). Through recent measurements
of CMBR/LSS, the parameters of the model have been constrained.

We have used the method of the continuous Bogoliubov coefficients to
calculate the GW energy spectrum for different
scales of energy and values of the $\alpha$-parameter of the model. 

Besides the fact that the model suffers a small deviation from the preferred values of the
spectral index $n_s$, the obtained spectra reveal a consistent picture
corresponding to a smooth transition between the inflationary and the radiation dominated epochs of the universe, and constitute a significant imprint of this modified generalised Chaplygin gas.

Finally, the strong variation of the high frequency range, is directly
related with the decrease in the maximum of the potential $a^{\prime\prime}/a$ (see Eq.~(\ref{6})) for $|\alpha|$ approaching $1$. This feature implies strong limits to the maximum frequency allowed in our model and which will be within the reach 
of future gravitational-waves detectors like BBO and DECIGO \cite{Lidsey97}, for the 
KHz range of frequencies. In fact, for the most consistent values of the $\alpha$-parameter, the spectrum
shows a frequency as low as in the Hz region.

Concluding, these results show that, for this model, the Hz-KHz frequency
range comes directly from the transition between the inflationary regime
and the radiation era, and addresses important issues about the limits of the GW energy spectrum.

Last but not least, we would like to highlight that one of the merits 
of the model we have presented for the transition from the inflationary era to the radiation dominated epoch is its relative simplicity.

\acknowledgments

The authors are grateful to O.~Bertolami and A.~Y.~Kamenshchik for very useful comments  on a previous version of the manuscript. M.B.L. is  supported by the Portuguese Agency Funda\c{c}\~{a}o para a Ci\^{e}ncia e Tecnologia through the fellowship SFRH/BPD/26542/2006. This research was supported by the grant FEDER-POCI/P/FIS/57547/2004.

\end{document}